\documentstyle[aps,prl,floats,graphicx,epsfig]{revtex}
\begin{document}
\draft
\twocolumn[\hsize\textwidth\columnwidth\hsize\csname @twocolumnfalse\endcsname

\title{Critical Velocity in  $^3$He-B
 Vibrating Wire Experiments as Analog of Vacuum Instability in a Slowly
Oscillating Electric Field.}
\author{A. Calogeracos$^{1,2}$ and G.E. Volovik$^{1,3}$}
\address{
$^{1}$Low Temperature Laboratory,
Helsinki University of Technology,
Otakaari 3A, 02150 Espoo, Finland\\
$^{2}$ NCA\ Research Associates,
 PO\ Box 61147, Maroussi 151 01,
Athens, Greece \\
$^{3}$ L.D. Landau Institute for Theoretical Physics,
Kosygin Str. 2, 117940 Moscow, Russia\\}
\date{\today}
\maketitle

\begin{abstract}

{The Lancaster experiments \cite{CCGMP} with a cylindrical wire moving in
superfluid $^3$He-B are discussed, where the measured critical velocity of
pair creation was much below the Landau critical velocity. The phenomenon is
shown to be analogous to the instability of the electron-positron vacuum in an
adiabatically alternating strong electric potential of both signs, where the
positive- and negative-root levels cross and thus the
instability treshold is twice less than in the conventional case of a
single static potential well.}
\end{abstract}

\

\pacs{PACS numbers: 11.90.+t , 67.57.-z , 74.60.Jg  }

\

] \narrowtext

\section{Introduction}

In superfluid Fermi system the pairs of quasiparticles are created by the
uniformly moving object, if its velocity exceeds the Landau critical velocity,
 $v_L=\Delta_0/p_F$. Here $p_F$ is the Fermi momentum,
$\Delta_0$ is the superfluid gap in bulk liquid. The critical velocity $v_L$
is also called the pair-breaking velocity; it marks the threshold of the
instability of the superfluid vacuum: breaking of Cooper pairs which
form the superfluid condensate. In the vacuum of the high energy physics
a similar situation can occur: (i) In a  strong electric field
\cite{Pomeranchuk,Gershtein,Greiner}; (ii) In a strong gravitational filed,
say, in the presence of the event horizon \cite{Zaumen}; (iii) If the
hypothetical object, which is external to the physical vacuum, moves with the
superluminal speed. Here we consider the pair creation in superfluid
$^3$He-B, which is analogous to the production of the electron-positron pairs
in  a  strong electric field.

Such experiments have been conducted in Lancaster \cite{CCGMP,CGPS}, where
a cylindrical wire vibrating in superfluid
$^3$He-B has been used as a moving object. It appeared that
  the measured critical velocity, at which a large  extra
dissipation of the wire was observed due to particle creations, was
essentially less than $v_L$. It was about 0.25$v_L$ independent of the
material and radius of the wire.

It was originally suggested
in \cite{CCGMP}, that such reduction has two origins:   a geometrical factor
$1/2$  results from the local enhancement of the velocity near the wire, while
the other reduction is related to the suppression of the gap
 in the vicinity of the surface of the wire, $\Delta<\Delta_0$. As a result,
the Landau criterium for the  filling of the surface bound states is
essentially smaller than $v_L$. However to provide
the momentum loss by the wire, the   quasiparticles must escape to infinity.
That is why the production of the  scattering states at subcritical velocity
has to be explained. This scenario was developed by Lambert
\cite{Lambert}, who showed that the adiabatic oscillation could do this job, if
the velocity amplitude of the wire exceeds some value, which was estimated as
$v^*=(1/5)v_L$.

We  develop further these arguments taking into account that in $^3$He-B (1)
the surface leads to the splitting of the gap, $\Delta_{\parallel}$ and
$\Delta_{\perp}$; and (2) the classical description of the bound state in the
surface layer should be substituted by the quantum mechanical one. We obtain
the modified value for $v^*$, which depends on the gap suppression.
Since the Bogoliubov-Nambu fermions in $^3$He-B are in many
respects similar to the Dirac electrons, we connect the
critical radiation of the quasiparticles by a slowly vibrating wire with the
instability of the electron-positron vacuum in the presence of a strong
electric field.  Our case corresponds to a slowly alternating electric
potential of both signs, which allows the electron-positron production at
essentially weaker field than in the conventional mechanism discussed by
Gershtein and Zeldovich \cite{Gershtein}. In this scenario the classical
positive- and negative-root solutions cross, which leads to the
particle-antiparticle production (see also discussion in Ref.
\cite{Jetzer}).  We constructed a simple time dependent potential for Dirac
electrons, which allows us to model the proposed scenario.

\section{Fermions in the vibrating wire.}
\subsection{Fermionic spectrum in $^3$He-B.}

In bulk superfluid $^3$He-B the fermionic spectrum is defined by the following
$4\times 4$ matrix Hamiltonian (Bogoliubov-Nambu Hamiltonian)
\cite{Vollhardt1990,Exotic}:
\begin{equation}
H({\bf p})= \beta M(p) +
 c  {\bf p}\cdot \vec\alpha ~, ~M(p)= v_F(p-p_F)~,~c={\Delta_0\over
p_F}~.
\label{H}
\end{equation}
Here $\beta$ and  $\vec\alpha$ are Dirac matrices, composed from the
$2\times 2$ Pauli matrices  $\vec \tau$  describing the Bogoliubov-Nambu
spin in particle-hole space and $2\times 2$ Pauli matrices $\vec \sigma$ for
conventional spin:
\begin{equation}
 \beta= \tau_3~~,~~\vec\alpha =\tau_1\vec\sigma   ~.
\label{DiracMatrices}
\end{equation}
 The energy spectrum is
\begin{equation}
E_\pm(p)=  \pm \sqrt { M^2(p) +
 c^2p^2}~.
\label{E}
\end{equation}
The quantity $c$ plays the part of speed of light, however as distinct from the
relativistic case the mass $M$ depends on the momentum
$p$. Since $v_F\gg c$ the minimum of the positive energy occurs not at
$p=0$ but
at $p=p_F$ with ${\rm min}~E_+(p)=\Delta_0$.

According to the Landau criterium, if the external body moves with the velocity
larger than $v_L={\rm min}~(E_+(p)/p)=c$ it will radiate the quasiparticles. As
distinct from the relativistic case, where the minimum is realized  at
$p\rightarrow \infty$, in $^3$He-B it occurs at
$p=p_F$.

In the reference frame of the body the energy spectrum is Doppler shifted:
\begin{eqnarray}
\nonumber H({\bf p})={\bf p}\cdot {\bf v}_s+ \beta M(p) +
 c  {\bf p}\cdot \vec\alpha ~,\\
E_\pm(p)= {\bf p}\cdot {\bf v}_s \pm \sqrt {
M^2(p) +
 c^2p^2}~.
\label{DopplerShifted}
\end{eqnarray}
where ${\bf v}_s$ is the superfluid velocity in the body frame. If
$v_s(\infty)>c$, the positive square-root continuum merges with the negative
square-root continuum and thus  the production  from the
vacuum of pairs of quasiparticles with momentum $p_F$ becomes possible.
Here we
discuss the situation when the particle production is possible even well below
the Landau criterium. It is the combined effect of (i) the enhancement of the
local superfluid velocity in the vicinity of the surface of the object;
(ii) the
decrease of the "speed of light" near the surface; and (iii) adiabatic
oscillation of the velocity of the body.

\subsection{Fermions in the surface layer.}

In the experimental situation \cite{CCGMP,CGPS} the external body moving in
$^3$He-B is the cylindrical wire of the radius $R$ from  2 to 50 $\mu$m,
which is much larger than the coherence length $\xi \sim v_F/\Delta_0$.  The
velocity of the wire performs the oscillating motion,
${\bf u}(t)=\hat {\bf x}u(t)$,
$u(t)=u_0\cos(\omega t)$, with frequency $\omega \sim 10^2-10^3$Hz, which
is much
smaller than the characteristic quasiparticle energy of order $\Delta_0$ and
thus the motion is extremely adiabatic. The presence of the moving external
object disturbs the vacuum state of the superfluid. First, the velocity
field is
modified by the moving wire. In the reference frame of the wire the superfluid
performs an ideal dipole flow around the wire:
\begin{equation}
{\bf v}_s({\bf r},t) =-{\bf u}(t) +  {R^2 \over  r^2} [2\hat {\bf r}( \hat {\bf
r}
\cdot {\bf u}(t)) -
{\bf u}(t)]~~, ~~r>R~~,
\label{LocalVelocity}
\end{equation}
where ${\bf r}=(x,y)$ is the 2D radius vector in the plane perpendicular to
the wire counted from the center of the wire; $\hat {\bf r}={\bf r}/r$. At two
lines at the surface of the wire the
superfluid velocity is twice larger than at infinity: ${\bf v}_s(\pm R
\hat {\bf y})=-2{\bf u}(t)$.

The second effect is that the order parameter (gap) is suppressed near the
surface of the wire in the layer of the thickness of the coherenece length
size. In $^3$He-B this suppresion is anisotropic, which leads
to the two "speeds of light"  in the region
$r-R \sim
\xi\sim v_F/\Delta_0$  (see Fig.\ref{Fig3}):
\begin{eqnarray}
\nonumber H  = \beta M(p) +
 (c_\parallel (\delta_{ij}-\hat n_i \hat n_j)+  c_\perp  \hat
n_i \hat n_j) p_i\alpha_j ~,\\
E_\pm(p)= {\bf p}\cdot {\bf v}_s \pm \sqrt {M^2(p)+c_\perp^2  (\hat{\bf n}\cdot
{\bf p})^2+c_\parallel^2  (\hat{\bf n}\times {\bf
p})^2} ~,
\label{HSurface}
\end{eqnarray}
where $c_\perp=\Delta_\perp/p_F$ and $c_\parallel=\Delta_\parallel/p_F$ are the
"speeds of light" along the normal
$\hat{\bf n}=\hat{\bf r}$ to the surface of the wire and parallel to the
surface
correspondingly. According to \cite{KSS}, where the  diffusive boundary
conditions were considered, the transverse speed of light is completely
suppressed, $c_{\perp}(r=R)=0$,   while $c_{\parallel}(r=R) \approx 0.4 c$ at
$T=0$.
Due to the suppression of the order parameter the surface layer serves as a
potential well for quasiparticles, which contains the bound states with
energies
below the gap\cite{Privorotskii} (see Fig.\ref{Fig3}).

\begin{figure}[!!!t]
\begin{center}
\leavevmode
\epsfig{file=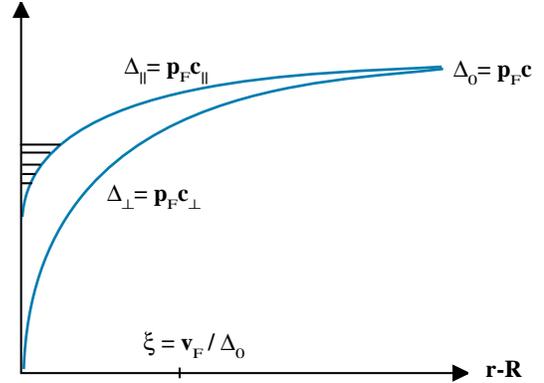,width=0.8\linewidth}
\caption[Fig3]
    {Schematic illustration of gaps, "speeds of light" and bound states
near the
surface of the wire.}
\label{Fig3}
\end{center}
\end{figure}
\section{Critical velocities and nucleation of quasiparticles.}

\subsection{Excitations of the bound states.}

Let us consider first the uniformly moving wire with constant velocity $u$. The
filling of the bound states can occur at a velocity smaller than the Landau
velocity $v_L$ for creation of the fermions in the continuous spectrum. This
velocity can be estimated from the Landau criterium for the classical spectrum
in Eq.(\ref{HSurface}) for the surface fermions. Since near the wall the
superfluid velocity is tangential, the Landau velocity for the nucleation
of the quasiparticles in the surface states is
$v_L^{\rm surface}={\rm min}~(E_+(p)/p_\parallel)=c_\parallel(r=R)$. The
minimum first occurs at $p_\parallel=p_F$ and $E_+=p_Fc_\parallel(r=R)$; note
that the transverse speed of light
$c_{\perp}(r)$ does not enter the criterium. Taking into account the
enhancement
of the superfluid velocity near the wall, one obtains that the negative energy
levels in the surface layer appear if the velocity  $u$ exceeds
\begin{equation}
v_0^*={1\over 2}c_\parallel(r=R)= v_L  {\Delta_\parallel(r=R)\over 2 \Delta_0}
\label{v_0^*}
\end{equation}
Here we used the Lambert notations for different critical velocities, see
Ref.\cite{Lambert} (in his
paper however he did not take into account the splitting of the gap and
assumed that $v_0^*$ is very small).

The situation does not
change if instead of the classical consideration of the energy spectrum in the
surface layer, one takes into account the quantization of the quasiparticle
motion along the normal to the wall. According to \cite{Privorotskii} the
quasicontinuum of the subgap bound states starts  above the energy
$p_Fc_\parallel(r=R)$ with
$p_\parallel\approx p_F$, which again gives $(1/2)c_\parallel(r=R)$ for the
Landau critical velocity for nucleation of the surface fermions.

Can the  negative energy
levels in the surface layer be filled by quasiparticles? For this it is
necessary to have the connection with the reservoir of quasiparticles. It
appears that this always occurs in our situation. The negative square-root
branch $E_-$ of the quasiparticles in Eq.(\ref{DopplerShifted}) is always
occupied.  When the velocity $u$ exceeds ${v_0^*}$, the energy of branch $E_-$
can be positive, while the energy of branch $E_+$ can be negative, so the
branches overlap and the quasiparticle from the filled branch $E_-$ can jump to
the empty level on $E_+$. Since momenta $p_x$ of these states are opposite,
this
can happen only if the momentum $p_x$ is not conserved, which is always the
case because of the surface roughness.

\subsection{Analog of Zel'dovich mechanism of positron nucleation.}

However when the surface Landau velocity is reached, the created
surface quasiparticles, which have zero energy in the wire reference frame,
cannot escape to infinity where the minimal energy of the scattering state is
$\Delta_0 -p_F u =\Delta_0[1-(1/2)(c_\parallel(r=R)/c)] >0$. For quasiparticles
to escape to infinity the velocity of wire must be essentailly higher. This
happens when  the lowest energy of the bound state $p_Fc_\parallel(r=R)
-2p_Fu_0$ merges with the continuum of the negative root states, whose upper
edge is at  $-\Delta_0 +p_Fu$. This gives the criterium for the  emission of
the quasihole, $u>v_1^*$
\begin{equation}
v_1^*={c+c_\parallel(r=R)\over 3}~.
\label{v_1^*}
\end{equation}
 This is equivalent to the
production of the positron  by the strong electrostatic potential well
discussed
by Zel'dovich, when the created electron fills the bound state, while the
positron is emitted to infinity.

\begin{figure}[!!!t]
\begin{center}
\leavevmode
\epsfig{file=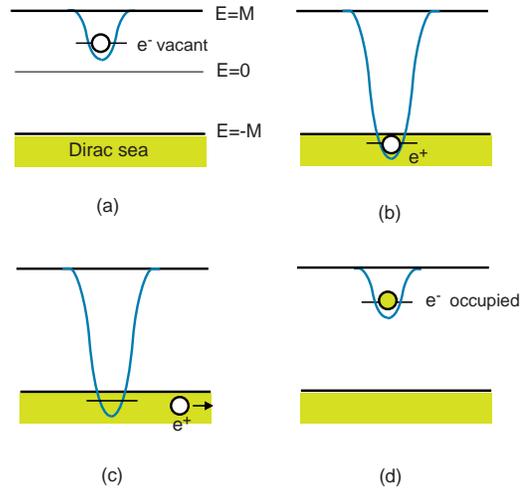,width=0.8\linewidth}
\caption[Fig4]
    {Zeldovich mechanism of positron creation.}
\label{Fig4}
\end{center}
\end{figure}

It may be helpful to remind the reader of the essential features of the
Zeldovich mechanism \cite{Gershtein} (see also \cite{Calogeracos} for a
detailed
review). Consider an electron-attractive potential with a vacant discrete level
(Fig.\ref{Fig4}(a)). Suppose that the potential adiabatically increases in
strength. The level will cross $E=0$ for some value $V_1$ of the potential
($V_1=\pi/2$ for a $\delta$-function potential). There is nothing critical
happening during the crossing. For some greater value $V_2$ the level crosses
$E=-M$ and thus merges with the negative energy continuum ($V_2=\pi $ for a
$\delta$-function potential). The original electron vacancy is now
interpreted as
the presence of the positron; and since the positron occupies a scattering
state
it can escape to infinity (Fig.\ref{Fig4}(c)). If the potential now becomes
weak
again we go back to the situation of a discrete energy level
(Fig.\ref{Fig4}(d)) which however now is electron-filled. The whole cycle
clearly conserves charge; however the positron escapes when the potential is
strong and the electron is observed when the potential returns to its original
weak value.

If the velocity of the object is kept constant, the  emission of
quasiparticles at $u>v_1^*$ will finally stop after all the negative levels
become occupied. Then the object will move without dissipation, but its
mass will
be larger due to the quasiparticles which occupied the negative energy bound
states. In the case of moving vortices in superfluids and
superconductors a similar enhancement of the mass due to the trapped
quasiparticles is the origin of the so-called Kopnin mass of the vortex (see
Ref.\cite{VortexMass}).

Thus for the uniformly moving object the dissipation is absent even if its
velocity exceeds
$v_1^*$, and nothing happens until the Landau velocity $v_L=c$ is reached, if
however the hydrodynamic instability does not develop earlier\cite{Parts}. The
source of this instability can be the following:  the filling of the bound
state
leads to increase of the normal component density and thus to the rearrangement
of the whole superflow pattern because of the mass conservation law (see
Ref.\cite{VortexMass} for the effect of the backflow due to the normal
component
in the vortex core). At some velocity the superflow pattern becomes unstable,
being unable to satisfy the mass conservation law. Such hydrodynamic
instability
leads usually to the formation of vortices by the moving object.

Eq.(\ref{v_1^*}) is analogous to the criterium obtained by Lambert
\cite{Lambert}, and it transforms to his result if
$c_\parallel(r=R)$ is neglected. However, in the real situation
$c_\parallel(r=R)/c$ is not small: it is close to unity for the specular
boundary
conditions, while for diffusive conditions it  is about
$c_\parallel(r=R)/c=0.4$
\cite{KSS}. Thus the most optimistic estimation gives
$v_1^*=0.47 v_L$ which is too large compared with the
experment, which shows that the supercritical dissipation starts at $\sim
0.25 v_L$.
Thus it appeared that the Zel'dovich mechanism in its simplest form is not
responsible for the supercritical behavior. The modification of this
mechanism is
required according to another scenario, also suggested by
Lambert\cite{Lambert}, who exploited the adiabatic oscillations of the wire.

\subsection{Radiation by adiabatically oscillating potential.}

The idea of this mechanism explores the fact that in the oscillating wire
$u=u_0 \cos(\omega t)$ the velocity changes sign after half a period. Let us
consider the case when the amplitude of the velocity $u_0 > {v_0^*}$ in
Eq.(\ref{v_0^*}). After the maximal velocity, say, $+u_0$ is reached, the bound
state with the energy $E_+= \Delta_0  c_\parallel(r=R)/c  - 2p_F {v_0^*}
=0$ will
be filled by the quasiparticle. Now, if the vibration of the wire is slow,
which
is the case since $\omega \ll \Delta_0$, then after the half of the period the
energy of this quasiparticle will become $E_+= \Delta_0 c_\parallel(r=R)/c +
2p_F {v_0^*}$.   We must compare this energy with the minimal energy of the
scattering states, which occurs for the opposite direction of the momentum:
$E_+(\rm min~scattering)=
\Delta_0 - p_F {v_0^*}$. So, if
\begin{equation}
  {v_0^*}> {v_L\over 5} ~,~~{\rm ~i.e.}~~~ {c_\parallel(r=R) \over c}>{2\over
5}~,
\label{ConditionOnCparallel}
\end{equation}
the continuum (conducting) energy band is achieved and the quasiparticles
will be
emitted by the vibrating wire. If, however $c_\parallel(r=R)  < (2/5) c$, then
the same mechanism starts to work at higher velocity, when
$u_0>(1/5) v_L$. The latter case corresponds to the Lambert result obtained
under
the assumption that the quantity ${v_0^*}$ is very small.
Thus the criterium for the radiation of the quasiparticles by the vibrating
wire
is
$u_0 > v^*$, with
\begin{eqnarray}
\nonumber  v^*={v_0^*}~,~~{\rm if}~~~ {v_0^*}>{v_L\over 5}\\
 v^*= {1\over 5}v_L ~,~~{\rm if}~~~ {v_0^*}<{v_L\over 5}~.
\label{v^*}
\end{eqnarray}

\begin{figure}[!!!t]
\begin{center}
\leavevmode
\epsfig{file=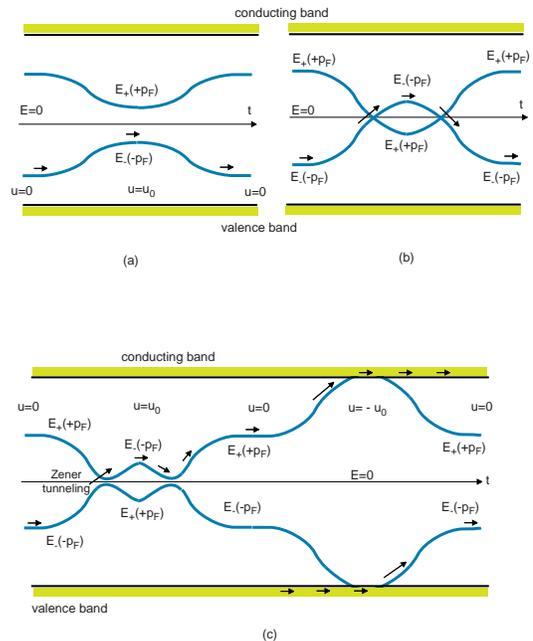,width=0.8\linewidth}
\caption[Fig1]
    {Time evolution of two
branches, $E_+(p_x=p_F)$ and $E_-(p_x=-p_F)$, of bound states. (a) Subcritical
regime. The half of the period is shown when the velocity increases till $u_0$
and then decreases till zero. (b) In the supercritical regime the two
branches cross each other, but the evolution of the levels does not change if
momentum $p_x$ is conserved.  (c) Level flow in the presence of mixing of
$+p_F$
and $-p_F$ states. The whole period of oscillations is shown, in which
"electron-positron" pair is created.}
\label{Fig1}
\end{center}
\end{figure}

The general scheme of the particle production at $u_0 > v^*$ is shown in
Fig.\ref{Fig1}.  In the supercritical
regime (b) in a time evolution the two branches, $E_+(p_x=p_F)$ and
$E_-(p_x=-p_F)$, of bound states cross each other, if the momentum
$p_x$ is conserved. In a real situation the surface
roughness mixes $+p_F$ and $-p_F$ states,
which leads to repulsion of levels. The time evolution of levels
and one of the trajectories of the quasiparticle  in supercritical regime are
shown in Fig. \ref{Fig1}(c)  during the whole period of oscillation. The
transition of the quasiparticle from the branch $E_-(-)$ to the branch $E_+(+)$
occurs either by scattering or Zener tunneling. During one cycle the particle
moves from the Dirac sea to the positive energy continuum via the bound states.
This  corresponds to the production of the electron-positron pair via the bound
states.

This mechanism is different
from the Zeldovich mechanism, in which the bound state energy touches the
continuum spectrum of the Dirac sea, the electron occupies the  bound
state, and the positron is emitted.  In our case the criticality occurs when
the bound state energy of the branch  $E_+$ reaches the zero energy and thus
touches the occupied bound states of the branches
$E_-$. In this process two particles in the scattering states are created
("electron" and "positron"), resulting in the production of the
momentum $2p_F$ from the vacuum. The level flow along two other
branches, $E_-(p_x=p_F)$ and $E_+(p_x=-p_F)$, is similar but is
shifted by half a period. As a result in this process an opposite momentum,
$-2p_F$, can be produced during a cycle.

\section{Analogy with fermion production in a strong electric field.}

Since close to the threshold velocity the relevant quasiparticle momentum
$p_x$ is maximal, $p_x=\pm p_F$,  the term ${\bf p}\cdot {\bf v}_s$ in
Eq.(\ref{DopplerShifted}) serves as the time like component of the 4-vector
electromagnetic potential: ${\bf p}\cdot {\bf v}_s=\pm
p_F v_{sx}(x,t)=eA_0(x,t)$. Here the sign of the momentum plays the part of the
electric charge. Thus we have a problem of Dirac particles in a strong electric
field. The above mechanism of particle creation requires 4 ingredients:

(1) Bound states.

(2) For the filling of the negative energy levels above
$v_0^*$ it is necessary to have the mirror image branch of quasiparticles with
opposite momentum (i.e. with opposite $e$).

(3) There should be the interaction which mixes the momenta $p_F$ and $-p_F$
and thus allows to change the sign $e$.

(4) The potential $A_0$ should be strong enough for the positive-root
and negative-root branches to cross.

(5) The potential $A_0$ should slowly oscillate in time. During
one cycle the positive-root and negative-root levels cross and then return to
their respective (positive/negative) continua.

That is why, in the mapping to the Dirac problem we would need the particles
with both negative and positive charges, which can transform to each other. One
possibility is to use instead of the  time-like component of the 4-vector
electromagnetic potential a time and space dependent mass term. In this case
the spectrum is symmetric, so that the positive and negative energy bound
states
can in principle approach each other \cite{Greiner}, in a similar
manner as in Fig.\ref{Fig1}(b).

\begin{figure}[!!!t]
\begin{center}
\leavevmode
\epsfig{file=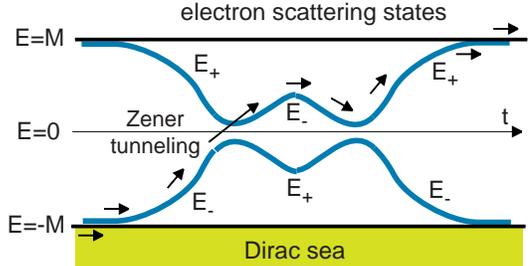,width=0.8\linewidth}
\caption[Fig2]
    {Spectral flow and pair production in the system of potentials
alternating in counterphase in supercritical regime.}
\label{Fig2}
\end{center}
\end{figure}

The other possibility is to have the conventional electromagnetic field $A_0$,
but in the form of two spatially separated potentials with the opposite sign of
$A_0$. In this case one has the required mirror image of states. This can be
modelled by the conventional Dirac Hamiltonian with potential
\begin{equation}
  A_0(x,t)=  U  \cos (\omega t) [\delta (x+a) - \delta(x-a)]~,
\label{A_0}
\end{equation}
Assuming that the Dirac mass $M=1$, and oscillations are adiabatic, $\omega
\ll 1$, one obtains the time dependent bound states energy levels
\begin{equation}
  E^2= \cos^2\lambda +   e^{-4ka}\sin^2\lambda~,~\lambda=   U  \cos
(\omega t)~,~k^2=1-E^2~.
\label{Solution}
\end{equation}
If $U $ exceeds the critical value $U_1=\pi/2$ the first (positive
energy) bound state crosses $E=0$. If the
$\delta$-potentials are well separated, $a\gg 1$, the time-dependent energy
levels are in Fig.\ref{Fig2}.
Here $E_+$ and $E_-$  denote the bound state levels in the right and in the
left
 $\delta$-function potential correspondingly.  The probability  of
nucleation of electron-positron pair is determined by the transition between
the $E_-$ and
$E_+$ branches. For  $U$ slightly above but not very close to $U_1=\pi/2$ one
obtains the result similar to that for the Landau-Zener tunneling problem
\cite{LandauLifshitz}, with the probability  for pair creation per one cycle
\begin{eqnarray}
2P(1-P)~,~P=\exp\left(-{2\sqrt{\pi} {\cal T}^2\over \omega
\sqrt{U-U_1}}\right)~,\\
{\cal T}^2=e^{-4a}\ll U-U_1 ~.
\label{Zener}
\end{eqnarray}
If $\omega$  is large
enough, the transition between the $E_-$ and $E_+$ states is given by the
matrix
element ${\cal T}$, while for small $\omega$  the process
is determined by Zener tunneling across the gap $2{\cal T}$ between the
repulsing levels.

The similar effect in nuclear physics would correspond to the case, different
from that suggested by Gershtein and Zeldovich. In their case the positron
production is possible during collision of two heavy bare nuclei with  the
total charge $Z$ greater then supercritical $Z_c$, at which the electron bound
state with energy $E=-M$ appears. This would correspond to the critical
strength  $U_2=\pi$   of  the $\delta$-function potential. In our case
the critical strength is
$U_1=\pi/2$. This means that we need essentially less total charge
$Z$, at which the negative energy  bound state for electron appears, $E_+<0$.
But in addition nearby one should have a similar hypothetical  collision of the
anti-nuclei, which produces the potential of the opposite sign. If the latter
contains the  bound state  with
$E_-=E_+$, the electron occupying this bound state can tunnel to the bound
state
of the positively charged nucleus. As a result the electron-positron pair will
appear after such collision.

\section{Discussion.}

According to the discussed scenario the observed critical velocity for
the pair nucleation by a vibrating wire, $v_0^*\approx 0.25 v_L$ \cite{CCGMP},
is determined by the bound states near the surface of the wire and
thus by the suppression of the parallel gap at the surface of the wire in
Eq.(\ref{v_0^*}). This gives an experimental estimation for the suppressed gap,
$\Delta_\parallel(r=R)\approx  0.5 \Delta_0$, which is comparable to the
theoretical estimation $\Delta_\parallel(r=R)\approx  0.4 \Delta_0$ \cite{KSS}.
This consistency provides the experimental evidence for the modified Zeldovich
mechanism of  pair creation in a strong field, in which the particles can be
created by the subcritical electric potential because of the level crossing.

The other objects, whose motion can be used to simulate the particle
production from the vacuum, are the topological objects, vortices and domain
walls. On the production of the momentum from the vacuum by the
moving vortex, caused by the axial anomaly phenomenon, see review
\cite{Lammi}. The quasiparticle production by the moving soliton in
superfluid $^3$He-A due to the combined effect of the Schwinger pair
production, event horizon and ergoregion is dicussed in Ref. \cite{Jacobson}.

\section{Acknowledgements}

One of us (AC) wishes to thank
the Low Temperature Laboratory of Helsinki University
of Technology for the hospitality and EU Human Capital and Mobility
Visitor Programme CHGECT94-0069 for its support.

\end{document}